\documentclass{ws-ijmpe}
\usepackage[super,compress]{cite}
\usepackage{breakurl}

\usepackage{nameref}
\usepackage{hyperref}
\hypersetup{
	colorlinks,
	citecolor=black,
	filecolor=black,
	linkcolor=black,
	urlcolor=black
}

\begin{document}

%\markboth{N. Yakoneko}{}

\catchline{}{}{}{}{}
%%%%%%%%%%%%%%%%%%%%%%%%%%%%%%%%%%%%%%%%%%%%%%%%%%%%%%%%%%%%%%%%%%%%

\title{
Induced surface and curvature tensions  equation of state of  hard spheres and its virial coefficients
%		\footnote{For the title,try not to use more than 3 lines. Typeset the title in 10 pt Times roman, uppercase and boldface.}
}

\author{Nazar S. Yakovenko
%	\footnote{Typeset names in 10~pt Times roman, uppercase. Use the footnote to indicate the present or permanent address of the author.}
}

\address{Department of Physics, Taras Shevchenko National University of Kyiv, 03022 Kiev, Ukraine\\
nsyakovenko@gmail.com}

%%\address{University Department, University Name, Address\\City, State ZIP/Zone, Country
%\footnote{State completely without abbreviations, the
%affiliation and mailing address, including country. Typeset in 8~pt
%Times italic.} \\
%%first\_author@university.edu}

\author{Kyrill A. Bugaev}

\address{Bogolyubov Institute for Theoretical Physics, Kyiv 03680, Ukraine\\
Department of Physics, Taras Shevchenko National University of Kyiv, 03022 Kiev, Ukraine\\
bugaev@th.physik.uni-frankfurt.de}

\author{Larissa V. Bravina}
\address{Department of Physics, University of Oslo, POB 1048 Blindern, N-0316 Oslo, Norway}

\author{Evgeny E.  Zabrodin}
\address{Department of Physics, University of Oslo, POB 1048 Blindern, N-0316 Oslo, Norway\\
Skobeltsyn Institute of Nuclear Physics, Moscow State University, 119899 Moscow, Russia}

\maketitle

\begin{history}
\received{Day Month Year}
\revised{Day Month Year}
%\accepted{Day Month Year}
%\comby{(xxxxxxxxxx)}
\end{history}

\begin{abstract}
Here we present new results obtained for the equation of state with induced surface and curvature tensions.
The explicit formulas for  the first five virial coefficients of system pressure and for  the induced surface and curvature tension coefficients are derived and  their possible applications are briefly discussed. 
\end{abstract}


\ccode{PACS numbers: 25.75.Ag, 24.10.Pa}

%\tableofcontents

\section{Introduction}
\label{sec:Intro}

During the last few years, the statistical mechanics of strongly interacting matter made a few steps in the direction of departing 
the framework of Van der Waals \cite{VDWeos} equation of state (EoS) and move towards more realistic EoS. 
A few years ago the concept of induced surface tension (IST) for the mixture of nuclear clusters of all sizes was suggested 
\cite{NY_IST0} in order to explain the mystery  of  why the statistical multifragmentation model \cite{NY_SMM}
which employes the eigen volume approximation instead of the excluded one works so well at low particle number densities. 
Then this concept was  successfully applied to the description of  experimental hadronic multiplicities measured in the collisions of heavy ions for the center-of-mass  collision energies  from $\sqrt{s_{NN}}=2.7$ GeV (lowest AGS BNL energy)  to  $\sqrt{s_{NN}}=2.76$ TeV (LHC CERN) \cite{IST1,IST2}. As an outcome of these efforts, the best description of all existing hadronic multiplicities  data 
with the fit quality $\chi^2/dof \simeq 1.1$ was achieved  \cite{IST1,IST2}.  These results were obtained using  only four different hard-core radii of hadrons, namely hard-core radii of pions $R_\pi$, kaons $R_K$, of other mesons $R_m$ and baryons $R_b$. In other words, having two additional global fitting parameters, i.e.  $R_\pi$ and $R_K$,
to the usual ones, i.e.  $R_m$ and  $R_b$, one could greatly improve the quality of the data description. 

 Very recently   the IST EoS with the realistic  multicomponent hard-core repulsion 
was applied  to model the mixture of hadrons and  light (anti)nuclei ((anti)deuterons, (anti)hyper-tritons, (anti)helium-3 
and (anti)helium-4) \cite{Grinyuk2020} and, in contrast to other approaches,  the IST EoS of Ref. \citen{Grinyuk2020}   used  the true classical second virial coefficients of light (anti)nuclei  and hadrons. 
In Ref. \citen{Grinyuk2020}   the IST EoS was applied  to describe  the multiplicities of hadrons and such nuclei  measured by the ALICE CERN  collaboration in Pb+Pb collisions at the center-of-mass collision energy $\sqrt{s_{NN}} =2.76$ TeV 
\cite{KAB_Ref1a,KAB_Ref1b,KAB_Ref1c}.
It is necessary to stress that this   approach has no analogs since it  allows one to 
obtain  an unprecedentedly high quality of description of 18 experimental data points with 3 fitting parameters  and to reach    
 $\chi^2_{tot}/dof \simeq 0.769$.

Also, the IST concept proved its worth in describing the properties of nuclear matter with very few adjustable parameters \cite{Ivanytskyi18}. It is remarkable that using only four adjustable parameters the EoS of Ref. \citen{Ivanytskyi18}
was able to simultaneously reproduce three major properties of nuclear matter, the value of incompressibility factor in the desired range
and the proton flow constraint \cite{Danielewicz} which alone consists of  eight independent mathematical conditions. 
Thus, with only 
four parameters the EoS of Ref.  \citen{Ivanytskyi18} obeys twelve conditions.  This is the highly nontrivial result since 
many EoS based on the relativistic mean-field approach are not able to obey the proton flow constraint \cite{Danielewicz}  having 10-15 adjustable parameters (see the compilation in Ref. \citen{Dutra14}). 

The main reason for such a success of the IST EoS is that, having a single additional parameter $\alpha=1.245$ compared to the one component Van der Waals EoS, it is able to reasonably well reproduce not only the second but also  the third and fourth virial coefficient of classical hard spheres \cite{IST2, QIST18}.  This approach  was further developed and refined for the 
systems with any number of hard-core radii  for the mixtures of  quantum gases of hard spheres \cite{QIST19} and the mixtures of  classical  hard spheres and hard discs \cite{yakovenko2019concept}. 

It is necessary also to mention that the quantum generalization \cite{NY_Vovch17} of the famous Carnahan-Starling EoS was suggested recently \cite{CSEoS}. Although this is an interesting and promising approach, but in our opinion, it requires further refinement and generalization to the mixtures of  quantum particles of  different hard-core radii. 

Despite these achievements of the IST EoS  one important element of IST concept  did not get proper attention yet. Namely 
an important  fact, that the concept of  IST and its generalization which includes the curvature tension (ISCT) \cite{yakovenko2019concept} allows one to quantify the influence of the dense thermal medium on the  effective 
excluded volume of particles and on their effective surface (see below),  was not discussed yet.  
One of the reasons for the absence of such a discussion is that  there were no simple analytical formulas for such an analysis.
Therefore, in this work we present the virial expansions not only for the pressure of hard spheres but also for the coefficients
of induced surface  and curvature tensions. Having such expansions up the fifth order one can study the density dependence 
of    the effective  excluded volumes of particles  and their effective surface. 
The present study is rather important,  since
the lack  of  information on the medium influence on the collective properties of particles and  their large  clusters,
probably,  is partly  responsible for the  absence of the microscopic theory of surface tension of classical liquids \cite{Frank}.

The work is organized as follows. In Sect. \ref{sec:ISCT EoS} we discuss the ISCT EoS for one-component systems. Sect. \ref{sec:Vir Exp} is devoted to deriving the virial coefficients for pressure, and for the coefficients
of induced surface  and curvature tensions of hard spheres. The effective excluded volume and effective surface of particles 
are briefly discussed in Sect. \ref{sec:Excl Vol}, while our conclusions are summarized in Sect. \ref{sec:Conclusions}.

\section{EoS of Induced Surface and Curvature Tensions}
\label{sec:ISCT EoS}

The ISCT EoS is a system of three equations for the pressure $p$, the induced surface tension coefficient $\Sigma$
and the induced curvature coefficient $K$ \cite{yakovenko2019concept}. 
For one sort of particles, this EoS  can be written as the following system 
\begin{align}\label{NY_Eq1}
%\label{vir exp general form}
p =& T  \phi \exp \left[ \frac{\mu}{T} - v \frac{p}{T} - s \frac{\Sigma}{T} -c \frac{K}{T}\right] ,
\\
\label{NY_Eq2}
\Sigma =& A T R \phi \exp \left[ \frac{\mu}{T} - v  \frac{p}{T} - s \alpha \frac{\Sigma}{T} -c \frac{K}{T} \right] 
,
\\
\label{NY_Eq3}
K =& B T R^2 \phi \exp \left[ \frac{\mu}{T} - v  \frac{p}{T} - s \alpha \frac{\Sigma}{T} -c \beta \frac{K}{T} \right] 
\,.
\end{align}
Here $\mu$ is a chemical potential of particles, while $\phi$,  $g$ $m$, $R$, $v$, $s$ and $c$ denote, respectively, one-particle thermal density, the degeneracy factor,   mass, hard-core radius, the eigen volume $v= \frac{4}{3} \pi R^3$, the eigen surface $s = \frac{3 v}{R}$  and eigen curvature $s = \frac{3 v}{R^2}$  of considered particles.  Their thermal density 
	\begin{equation}
	\label{NY_Eq4}
	\phi (T)  = g \int\limits \frac{dk ^3 }{(2\pi\hbar)^3} e^{-\frac{\sqrt{k^2+m^2}}{T}} ,
	\end{equation}
is given in the Boltzmann approximation. 
The dimensionless parameters $\alpha > 1$ and $\beta>1$ account for the high-density terms of virial expansion and allow us to go beyond the second virial approximation. In principle, $\alpha$ or $\beta$ can be a regular functions of $T$ and $\mu$, but, for the sake of simplicity,  they are fixed to be constants. The parameters $A$ and $B$  are introduced to evaluate the contribution of induced surface $\Sigma$ and curvature $K$ tensions more accurately.

\section{Virial Expansion Analysis}
\label{sec:Vir Exp}

The  virial expansions of functions  $p$, $\Sigma$ and $K$ is helpful since even at high particle number  densities  they  can be used as an initial  approximation to find a solution of the system  (\ref{NY_Eq1})-(\ref{NY_Eq3}). For one sort of particles from  the  ISCT EoS   (\ref{NY_Eq1})-(\ref{NY_Eq3})  one can obtain the useful  relations:
\begin{align}
\label{rel betw p, Sigm, K}
\Sigma = A p R \exp \left[-s (\alpha-1) \frac{\Sigma}{T}\right]
\,,\qquad
K = \frac{B}{A} \Sigma R \exp \left[-c (\beta-1) \frac{K}{T}\right]
\,. 
\end{align}
Writing the  pressure, induced surface, and curvature tensions coefficients  
 in a Taylor series in powers of the particle number density
 $\rho=(\frac{\partial p}{\partial \mu})_T$ and assuming that $B = 1 - A$ \cite{yakovenko2019concept} one obtains
\begin{align}
%\label{vir exp general form}
\label{Eq_Vir_p}
p &= T \left(a_1 \rho +  a_2 \rho^2 + a_3 \rho^3+ a_4 \rho^4 + a_5 \rho^5 \right) ,
\\
\label{Eq_Vir_S}
\Sigma &= A R T \left(b_1 \rho +  b_2 \rho^2 + b_3 \rho^3+ b_4 \rho^4 + b_5 \rho^5 \right) ,
\\
\label{Eq_Vir_K}
K &= (1-A) R^2 T \left(c_1 \rho +  c_2 \rho^2 + c_3 \rho^3+ c_4 \rho^4 + c_5 \rho^5 \right) .
\end{align}
Substituting these expressions into Eqs.   (\ref{NY_Eq1}) and (\ref{rel betw p, Sigm, K}), one can get the following
expressions for the virial coefficients of pressure
\begin{align}
		a_1 =& 1  	\,,  \\
		 \quad  a_2 =& 4 v  
		\,, \\
		\label{Eq_a3}
		a_3 =& 16 v^2-6 v \left(\gamma +\delta  \beta _p\right)  \,, \\
		\label{Eq_a4}
		a_4 =& 64 v^3 - 72 v^2 \left(\gamma +\delta  \beta _p\right)+\frac{9}{2} v \left(3 \gamma ^2+\delta  (4 \gamma +3 \delta ) \beta _p\right)   
		\,, \\
		\label{Eq_a5}
		a_5 =& 256 v^4  -576 v^3 \left(\gamma +\delta  \beta _p\right)+v^2 \left(288 \gamma ^2+216 \delta  (2 \gamma +\delta ) \beta _p+72 \delta ^2 \beta _p^2 \right)  \nonumber
		\\
		-& 2 v \left(16 \gamma ^3+\left(24 \gamma ^2 \delta +27 \gamma  \delta ^2+16 \delta ^3\right) \beta _p\right) 
		\,,
\end{align}
where for convenience  we used the following notations 
\begin{align}
\label{gamma, delta def}
	\gamma = 3 (\alpha -1) (1-\beta_p) v 
	\,, \quad
	\delta =3 \beta_p (\beta -1) v 
	\,, \quad \text{where} \quad
	\beta_p = 1 - A
	\,,
\end{align}
Similar expressions  were found for the coefficients $\{ b_k\}$ and $\{ c_k\}$, but they are given in \ref{appendix: Vir coeff}. 

\begin{figure}[th]
	
	\centerline{
	\includegraphics[width=0.6\linewidth]{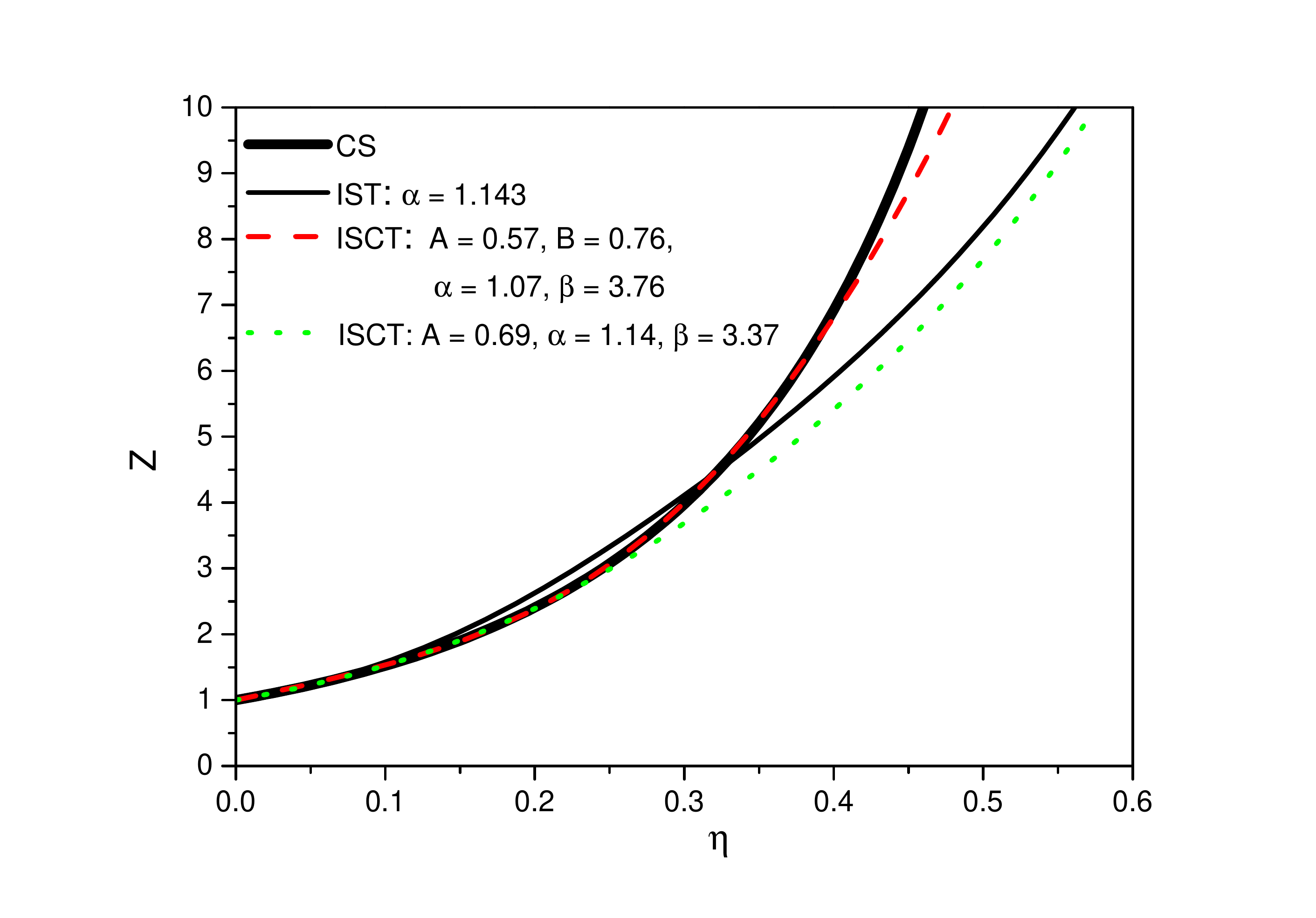}
		\hspace*{-11.2mm}
		\includegraphics[width=0.6\linewidth]{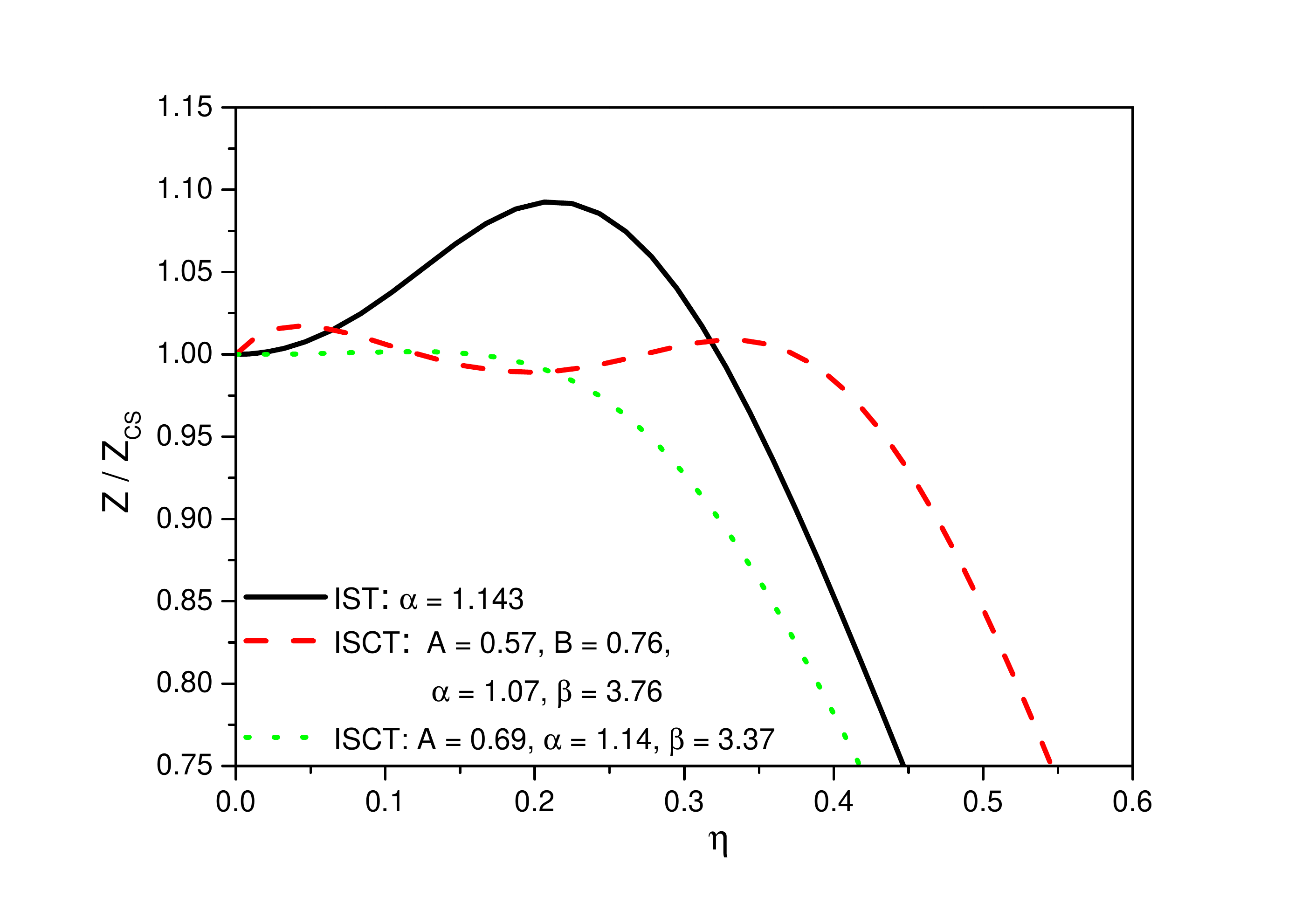}
}

	\caption{ {\bf Left panel.} Comparison of the compressibility factors $Z= \frac{p}{\rho T}$ of the CS EoS (solid thick curve) with the one-component IST EoS with best-fit parameters on interval of packing fraction $\eta \in [0.; 0.4]$ (solid thin curve), the ISCT EoS with best-fit parameters on the same interval (dashed curve) and  the  ISCT EoS which exactly reproduces the five virial coefficients of the CS EoS (dotted curve). {\bf Right panel.} Same as in the left panel, but for the ratios of the compressibility factors $Z/Z_{CS}$.}
	\label{NY_Fig1}
\end{figure}

Having   three parameters $\gamma$, $\delta$ and $\beta_p$ 
one can exactly reproduce five virial coefficients of the Carnahan-Starling EoS \cite{CSEoS} for hard spheres
or the Barrio-Solana  EoS \cite{BSEoS} or  their numeric representations taken from Refs. \citen{MC_for_HS}, \citen{MC_for_HD}.
For example,  the  virial expansion for the compressibility factor $Z_{CS}$ of the Carnahan-Starling EoS \cite{CSEoS} is 
\begin{align}
\label{ZCS vir exp}
Z_{CS} =& \frac{p}{\rho T} = \frac{1+\eta +\eta^2 -\eta ^3}{(1-\eta )^3} 
\\
 \approx& 1 + 4\eta + 10\eta^2 + 18\eta^3 + 28\eta^4 + 40\eta^5 + 54\eta^6 + 70\eta^7
\,,
\end{align}
where $\eta = \rho v$ is a packing fraction, can be reproduced by substituting  the coefficients  $a_3^{CS} = 10v^2$, $a_4^{CS} = 18v^3$ and $a_5^{CS}= 28v^4$ into the  left hand side of Eqs.  (\ref{Eq_a3})-(\ref{Eq_a5}). Then one can analytically obtain the values of  the coefficients  $\gamma$, $\delta$ and $\beta_p$.  However,  exact expressions are rather involved and, hence,  we give only the approximate numeric solutions:
\begin{align}\label{NY_Eq17}
\gamma = 0.302706 v,\quad \delta = 2.22698 v,\quad \beta_p= 0.313112
\,;\\
\label{NY_Eq18}
\gamma =-4.79862 v, \quad \delta =2.75922 v, \quad \beta_p=2.10154
\,.
\end{align}
Apparently, the solution (\ref{NY_Eq18})   with  $\beta_p > 1$ is unphysical since it corresponds to   negative value of the coefficient $A= 1-\beta_p$ and, consequently, to negative values of the surface tension coefficient $\Sigma$. 
From the solution (\ref{NY_Eq17}) one can find  the corresponding parameters  which enter the system (\ref{NY_Eq1})-(\ref{NY_Eq3}) as  $A \simeq 0.69, \alpha \simeq  1.14, \beta \simeq  3.37$ (see the dotted curve in Fig. \ref{NY_Fig1}). 

To demonstrate the advantage of the ISCT EoS in Fig. \ref{NY_Fig1} 
we compare the CS EoS with the  IST and ISCT EoS for which the coefficients  $\alpha$, $\beta$, $A$ and $B$ parameters 
were found to  reproduce the CS EoS on an interval of packing fraction $\eta$ up to $0.4$ \cite{yakovenko2019concept}.
Also  in Fig. \ref{NY_Fig1} we show the ISCT EoS  for the parameters which correspond to the solution (\ref{NY_Eq17}) which exactly reproduces  five virial coefficients of the CS EoS. For definiteness  we made calculations for the nucleon-like hard spheres, i.e. for $m =938.9$ MeV and  for  $g = 4$. For the hard-core radius of nucleons  we used  the typical value $R=0.39$ fm \cite{Ivanytskyi18}. As one can see from Fig. \ref{NY_Fig1} the ISCT EoS is able to accurately reproduce 
the compressibility $Z_{CS}$  (\ref{ZCS vir exp}) up to $\eta \simeq 0.42$, while the virial expansion (\ref{Eq_Vir_p})
provides a good description of $Z_{CS}$ up to $\eta \simeq 0.25$. 

To reproduce five virial coefficients of the Monte-Carlo simulations for gas of hard-spheres\cite{MC_for_HS}
\begin{equation}
\label{Z mc for hard spheres}
Z_{MC} = 1 + 4\eta + 10\eta^2 + 18.36\eta^3 + 28.23\eta^4 + 39.74\eta^5 + 53.5\eta^6 + 70.8\eta^7
\end{equation}
one should equate  $a_3 = 10v^2$, $a_4 = 18.36v^3$ and $a_5= 28.23v^4$ to the corresponding coefficients given by  Eqs.  (\ref{Eq_a3})-(\ref{Eq_a5}). Then the  solution with  positive values of $\Sigma$ is 
\begin{align}\label{NY_Eq20}
\gamma = 0.316292 v,\quad \delta = 2.28784 v,\quad \beta_p= 0.298844
%%\,;\\
%%\gamma =-4.88903 v, \quad \delta =2.79143 v, \quad \beta_p=2.10968
\,.
\end{align}
Comparing the solutions (\ref{NY_Eq17}) and (\ref{NY_Eq20}) one can see that the difference between the coefficients  is a few percent only.

\section{Effective Excluded Volume and Effective Surface of Hard Spheres}
\label{sec:Excl Vol}

\begin{figure}[th]
	
	\centerline{\hspace*{2.8mm}
	\includegraphics[width=0.56\linewidth]{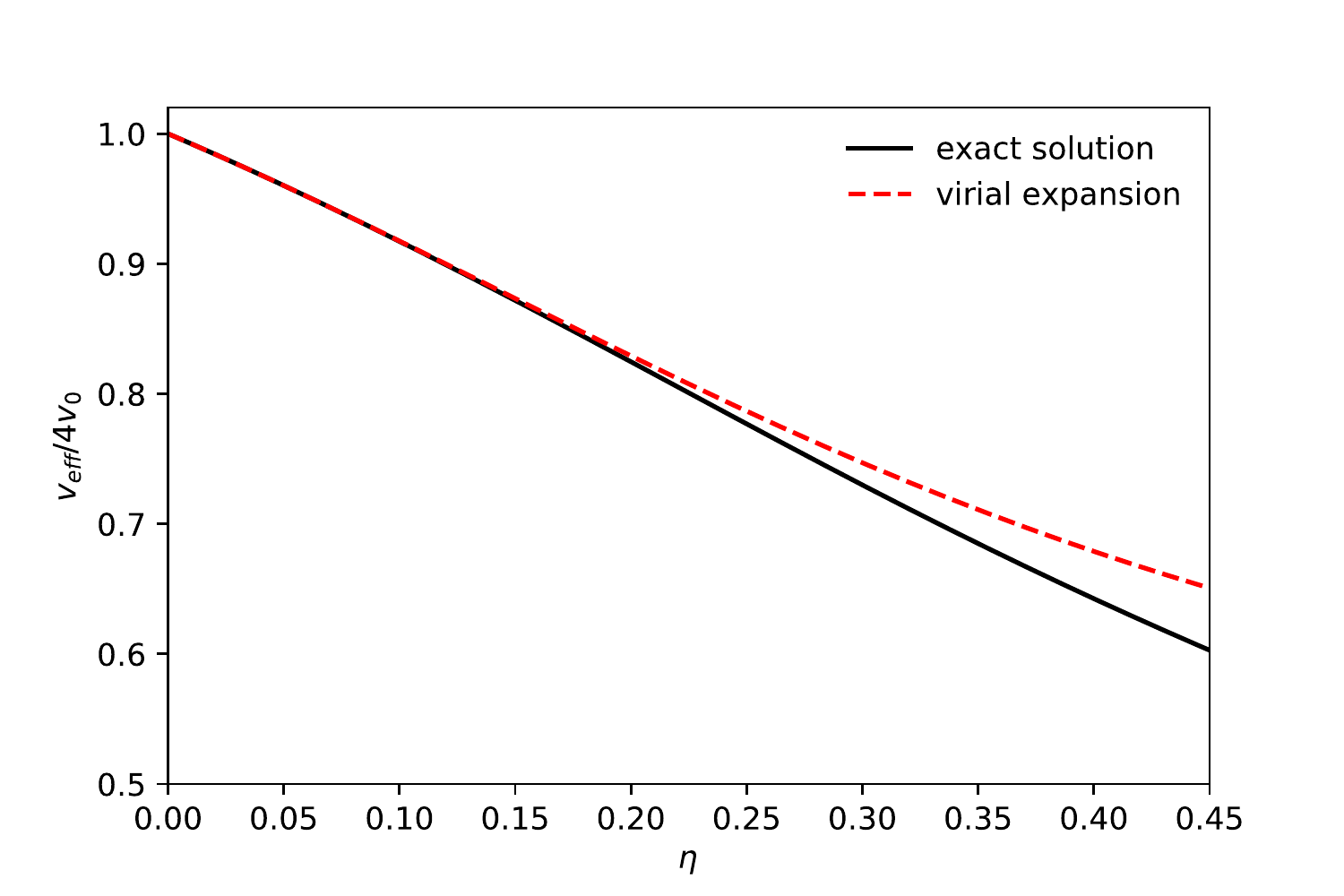}
		\hspace*{-7.mm}
		\includegraphics[width=0.56\linewidth]{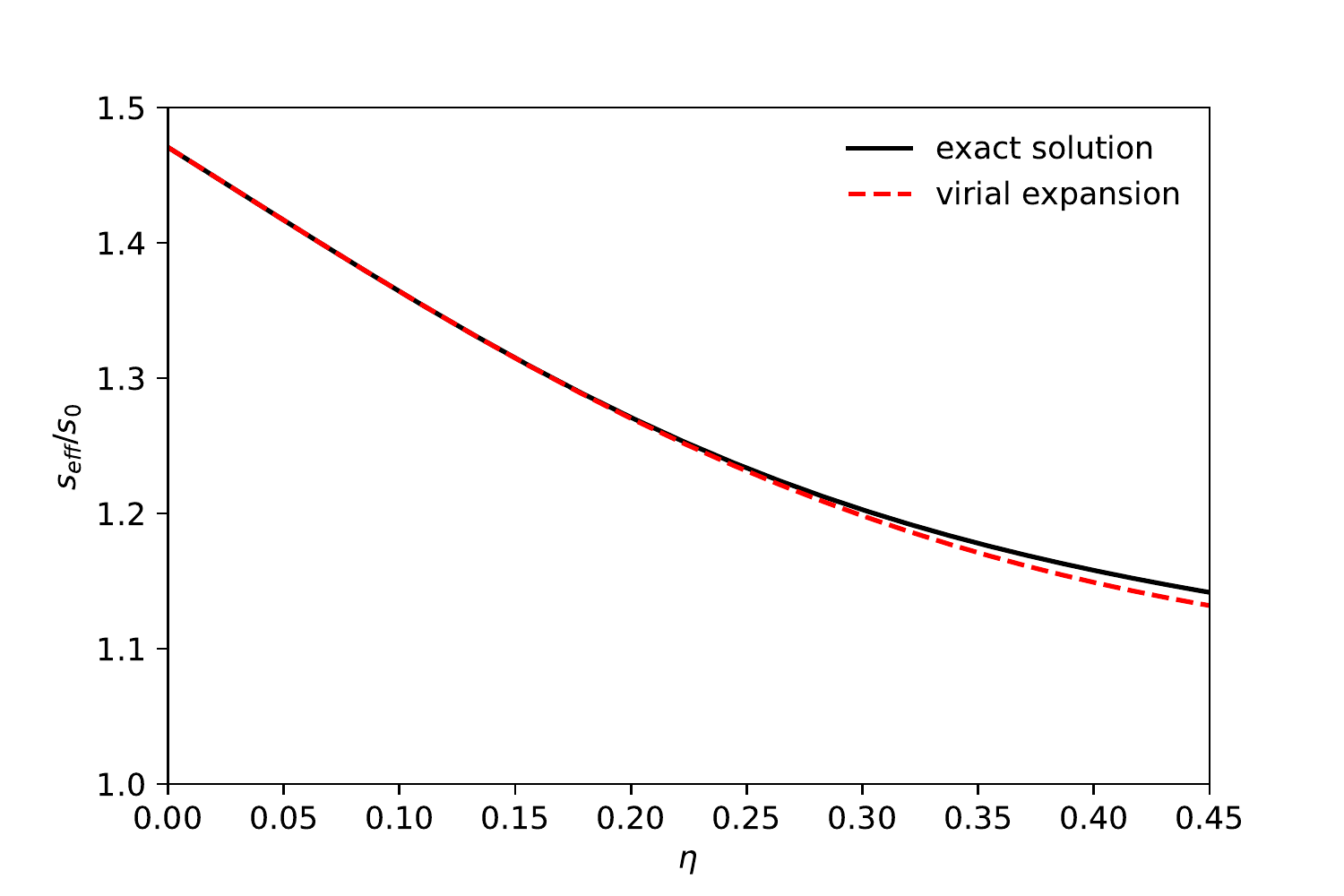}
}

	\caption{ {\bf Left panel.} Comparison of  $\eta$ dependence of  the effective excluded volume $v_{eff}$  (\ref{NY_Eq21}) found from  the ISCT EoS (solid curve) and from the virial expansion (\ref{Eq_Vir_p})-(\ref{Eq_Vir_K}) for the solution (\ref{NY_Eq17})
	(dashed curve).
{\bf Right panel.} Same as in the left panel, but for the effective surface $s_{eff}$  (\ref{NY_Eq21}). }
	\label{NY_Fig2}
\end{figure}

Here we define the effective excluded volume $v_{eff}$ and effective surface $s_{eff}$ of  particles 
\begin{eqnarray}
 \frac{v_{eff}}{4 v} = \frac{v  p  +  s  \Sigma +   c K}{4 v p} \,, \quad  \frac{s_{eff}}{s} = 1+ \frac{   c K}{ s \Sigma} ,
\label{NY_Eq21}  
\end{eqnarray}
as the density-dependent quantities.    Fig. \ref{NY_Fig2} explains the reason of why the ISCT EoS is more elaborate 
than the Van der Waals \cite{VDWeos} and the IST EoS \cite{Ivanytskyi18}. Indeed,  the left panel of  Fig. \ref{NY_Fig2} demonstrates that in the ISCT EoS the effective excluded volume  $v_{eff}$ sizably decreases if  the particle number density grows, while in  the Van der Waals  the excluded volume is constant. Similarly, as one can see from the right panel of
 Fig. \ref{NY_Fig2},  the effective surface of particles $s_{eff}$ also decreases, if the particle number density grows,
 but in the IST EoS \cite{Ivanytskyi18} the surface of particles is constant. 

The density  dependence of the  effective excluded volume $v_{eff}$ can be used to develop the quantum EoS similar to the 
one suggested in Ref.  \citen{NY_Vovch17}, while the density-dependent effective surface of particles $s_{eff}$ may be useful to
develop more realistic EoS of nuclear matter in which the total surface tension (being the sum of eigen surface tension and the induced one)  of large nuclear fragments vanishes at the critical endpoint, as it should be. 

\section{Conclusions}
\label{sec:Conclusions}

In this work, we present the virial coefficients  of the  recently developed  EoS based on the  ISCT  concept \cite{yakovenko2019concept}.  Besides the induced surface tension coefficient $\Sigma$  this concept   takes into account the induced curvature tension coefficient $K$. Both tensions are generated by the hard-core repulsion of hard spheres. 
Similarly to the pressure $p$, the functions  $\Sigma$ and $K$ also can be expanded into the virial expansion with the coefficients 
which are $T$-independent.   The explicit expressions for the first five virial coefficients of  $p$,  $\Sigma$ and $K$ 
are presented here. They can be used for simple analytical estimates  and  development of more elaborate EoS 
of nuclear and neutron matters.

\vspace*{-4.4mm}

\section*{Acknowledgments}
N.S.Ya.,  K.A.B., and L.V.B.  thank the Norwegian Agency for International Cooperation and Quality Enhancement in Higher Education for financial support, grant 150400-212051-120000 "CPEA-LT-2016/10094 From Strong Interacting Matter to Dark Matter". 
The work of K.A.B. was supported in part by  the Program of Fundamental
Research of the Department of Physics and Astronomy of the National
Academy of Sciences of Ukraine (project No. 0117U000240).
The work of L.V.B. and E.E.Z. was supported by the Norwegian 
Research Council (NFR) under grant No. 255253/F50 - CERN Heavy Ion 
Theory. 
K.A.B. is thankful  to the  COST Action CA15213 for supporting his networking.

\appendix
\section{Virial expansion coefficients}
\label{appendix: Vir coeff}

For the  one-component ISCT EoS  (\ref{NY_Eq1})-(\ref{NY_Eq3}) we also  found the virial coefficients for the surface tension coefficient $\Sigma$ (\ref{Eq_Vir_S})
\begin{align}
	b_1 =&  1  \,, \quad b_2 =  4 v-\gamma   
	\,, \\
	b_3 =&  16 v^2 + \frac{3 \gamma ^2}{2}+v \left(-14 \gamma -6 \delta  \beta _p\right)
	\,, \\
	b_4 =& 64 v^3   -\frac{8 \gamma ^3}{3}+v^2 \left(-120 \gamma -72 \delta  \beta _p \right)+v \left[\frac{87 \gamma ^2}{2}+\left(30 \gamma  \delta +\frac{27 \delta ^2}{2}\right) \beta _p\right]
	\,, \\
	b_5 =&  256 v^4 + \frac{125 \gamma ^4}{24}+v^3 \left(-832 \gamma -576 \delta  \beta _p\right)+v^2 (624 \gamma ^2+24 \delta  (26 \gamma +9 \delta ) \beta _p  \nonumber
	\\
	+& 72 \delta ^2 \beta _p^{2} )+v \left[\left(-111 \gamma ^2 \delta -81 \gamma  \delta ^2-32 \delta ^3\right) \beta _p-\frac{386 \gamma ^3}{3}\right]
	\,, 
\end{align}

and the virial coefficients for the curvature tension coefficient $K$ (\ref{Eq_Vir_K})
\begin{align}
	c_1 =&  1 \,, \quad 	c_2 =  4 v  -\gamma -\delta
	\,, \\
	c_3 =&  16 v^2 + \frac{3 \gamma ^2}{2}+2 \gamma  \delta +\frac{3 \delta ^2}{2}+v \left(-14 \gamma -8 \delta -6 \delta  \beta _p\right) 
	\,, \\
	c_4 =&  64 v^3  -\frac{8 \gamma ^3}{3}-4 \gamma ^2 \delta -\frac{9 \gamma  \delta ^2}{2}-\frac{8 \delta ^3}{3}+v^2 \left(-24 (5 \gamma +2 \delta )-72 \delta  \beta _p\right) \nonumber
	\\ 
	+& v \left[\frac{87 \gamma ^2}{2}+36 \gamma  \delta +18 \delta ^2+\left(30 \gamma  \delta +\frac{51 \delta ^2}{2}\right) \beta _p\right]
	\,, \\
	c_5 =&  256 v^4  + \frac{1}{24} \left(125 \gamma ^4+200 \gamma ^3 \delta +270 \gamma ^2 \delta ^2+256 \gamma  \delta ^3+125 \delta ^4\right) \nonumber
	\\
	+& v^3 \left(-64 (13 \gamma +4 \delta )-576 \delta  \beta _p\right)+v^2 (48 \left(13 \gamma ^2+8 \gamma  \delta +3 \delta ^2\right)  \nonumber
	\\
	+& 24 \delta  (26 \gamma +17 \delta ) \beta _p + 72 \delta ^2 \beta _p^{2})+v \left[\frac{1}{3} (-386 \gamma ^3-381 \gamma ^2 \delta -297 \gamma  \delta ^2 \right.  \nonumber
	\\
	-& 128 \delta ^{3})+\left(-111 \gamma ^2 \delta -153 \gamma  \delta ^2-86 \delta ^3\right) \beta _p \biggl]
	\,.
\end{align}

% \bibliographystyle{ws-ijmpe}
% \bibliography{bibliography}		% Produces the bibliography via BibTeX.

\end{document}